\documentclass[aps,prd,onecolumn,showpacs,showkeys,amsmath,amssymb]{revtex4}
\usepackage{graphicx}
\usepackage{dcolumn}
\usepackage{amsmath}
\usepackage{bm}
\begin{document}
\title{Decuplet contribution to the meson-baryon scattering lengths}
\author{Yan-Rui Liu}\email{yrliu@pku.edu.cn}
\affiliation{Department of Physics, Peking University, Beijing
100871, China}
\author{Shi-Lin Zhu}
\email{zhusl@phy.pku.edu.cn} \affiliation{Department of Physics,
Peking University, Beijing 100871, China}

\begin{abstract}

We calculate decuplet contributions to the s-wave pseudoscalar
meson octet-baryon scattering lengths to the third order in heavy
baryon chiral perturbation theory (HB$\chi$PT). Using experimental
pion-nucleon and kaon-nucleon scattering lengths as inputs, we
determine low-energy constants and predict other meson-baryon
scattering lengths. Numerically we consider three cases: (1) the
case with only baryon octet contributions; (2) with decuplet
contributions and (3) in the large $N_c$ limit. Hopefully, the
analytical expressions and the predictions are helpful to future
investigations of the meson-baryon scattering lengths.

\end{abstract}



\maketitle

\pagenumbering{arabic}

\section{Introduction}\label{sec1}

Chiral perturbation theory involving only pseudoscalar mesons is
expanded with $p/\Lambda_\chi$ where $p$ represents the meson mass
or momentum, and $\Lambda_\chi\sim 1\mathrm{GeV}$ is the scale of
chiral symmetry breaking. When ground baryons are incorporated in
the Lagrangian, the chiral expansion is problematic because of
terms like $M_0/\Lambda_\chi \sim 1$, where $M_0$ is the baryon
mass in the chiral limit. This problem is overcome in HB$\chi$PT
\cite{hbchpt,hbchptD,review} by going to extremely
non-relativistic limit. Now one makes the dual expansion of
$p/\Lambda_\chi$ and $p/M_0$ simultaneously, where $p$ also
represents the small residue momentum of baryons in the
non-relativistic limit.

In low energy processes, decuplet baryon contributions may be
important. Firstly, the mass difference between decuplet and octet
baryons $\delta=294\mathrm{MeV}$ is not large. Furthermore, this
value vanishes in the large $N_c$ limit \cite{nc,ncWitten}.
Secondly, the coupling constant of decuplet and octet baryons with
pseudoscalar mesons is large. Thus, the inclusion of these states
may cancel some intermediate octet contributions. In fact, decuplet
contributions partially cancels the large octet contribution in
baryon axial currents \cite{hbchptD}.

One of the systematic approaches to include decuplet baryons in
HB$\chi$PT is the small scale expansion (SSE)\cite{SSE}. Within this
counting scheme, the meson masses, all the momenta and $\delta$ are
all of order ${\cal O}(\epsilon)$. This formalism was widely used to
study processes involving explicit $J=\frac32$ fields
\cite{ffn,ffn2,dnff,pnd,char,axial,axialD,vec}. Besides SSE, an
alternative counting scheme was proposed in \cite{DE,DE2,DE3}.

For the elastic scattering of the pseudoscalar meson and
octet-baryon, the scattering length $a_{PB}$ is an important
observable, which is related to the threshold $T$-matrix by $T_{P
B}=4\pi(1+{m_P\over M_B}) a_{PB}$. HB$\chi$PT provides a
model-independent approach to calculate this threshold parameter.
Chiral corrections to pion-nucleon scattering lengths were first
investigated in two-flavor HB$\chi$PT in \cite{pnsl,pnsl2}.
Intermediate $\Delta$ corrections to them can be found in
\cite{pnd}.

For the other meson-baryon interactions, one has to work in the
SU(3) framework. Now the convergence of the chiral expansion has to
be investigated channel by channel because of the large mass $m_K$
or $m_\eta$. In \cite{KN}, the s-wave kaon-nucleon scattering
lengths were calculated to ${\cal O}(p^3)$ in SU(3) HB$\chi$PT. We
calculated chiral corrections to octet-meson octet-baryon scattering
lengths to the third order \cite{MB}. In the present work, we will
consider the decuplet baryon contributions to the threshold
meson-baryon amplitudes to ${\cal O}(\epsilon^3)$ in SSE in SU(3)
HB$\chi$PT.

In the previous calculations \cite{KN,MB}, the counter-term
contributions at ${\cal O}(p^3)$ were assumed to be much smaller
than the loop contributions. This rather naive assumption is an
extension of the SU(2) case \cite{pnsl} where the counter-terms were
estimated with resonance saturation method and found to be small.
The assumption was used partly because the complete third order
meson-baryon chiral Largrangians were unknown. Recently, the
complete and minimal Lorentz invariant SU(3) chiral Lagrangians were
composed to ${\cal O}(p^3)$ \cite{Lp3,Lp32,Lp33}. One needs to
consider the counter-term contributions now.

In the following section, we collect the basic definitions and
Lagrangians. We present decuplet contributions to the threshold
$T$-matrices in Sec. \ref{sec3} and the counter-terms for the
third-order $T$-matrices Sec. \ref{sec4}. Then we determine the
low-energy constants (LECs) in Sec. \ref{sec5} by considering the
counter-term contributions. The final section is our numerical
results and discussions.

\section{Lagrangians}\label{sec2}

The Lagrangian of HB$\chi$PT with octet baryons has the form
\begin{equation}\label{lagr}
{\cal L}={\cal L}_{\phi\phi}+{\cal L}_{\phi B},
\end{equation}
where $\phi$ represents the pseudoscalar meson octet and $B$
represents the baryon octet. The purely mesonic part ${\cal
L}_{\phi\phi}$ incorporates even chiral order terms while ${\cal
L}_{\phi B}$ starts from ${\cal O}(p)$. When decuplet baryons are
incorporated into the system, an additional part ${\cal L}_{\phi
BT}$ is introduced in Eq. (\ref{lagr}) where $T$ represents the
baryon decuplet. The lowest order Lagrangians of the three parts are
\begin{equation}
{\cal L}^{(2)}_{\phi\phi}=f^2 {\rm tr}(u_\mu
u^\mu+\frac{\chi_+}{4}),
\end{equation}
\begin{equation}
{\cal L}^{(1)}_{\phi B}={\rm tr}(\overline{B}(i\partial_0
B+[\Gamma_0,B]))-D {\rm tr}(\overline{B}\{\vec\sigma\cdot\vec
u,B\})-F {\rm tr}(\overline{B}[\vec\sigma\cdot \vec u,B]),
\end{equation}
\begin{equation}
{\cal L}^{(1)}_{\phi BT}=-\overline{T}^\mu(i{\cal D}_0-\delta)T_\mu
+{\cal C}(\overline{T}^\mu u_\mu B+\overline{B}u_\mu T^\mu) -{\cal
H} \overline{T}^\mu \vec\sigma\cdot \vec u T_\mu,
\end{equation}
where $\delta$ is the decuplet and octet baryon mass difference in
the chiral limit and the common notations read
\begin{equation}
\Gamma_\mu = {i\over 2} [\xi^\dagger, \partial_\mu\xi],\qquad
u_\mu={i\over 2} \{\xi^\dagger, \partial_\mu \xi\},\qquad \xi =
\exp(i \phi/2f),
\end{equation}
\begin{equation}
\chi_\pm = \xi^\dagger\chi\xi^\dagger\pm\xi\chi\xi,\qquad
\chi=\mathrm{diag}(m_\pi^2,\, m_\pi^2,\, 2m_K^2-m_\pi^2),
\end{equation}
\begin{eqnarray}
\phi=\sqrt2\left(
\begin{array}{ccc}
\frac{\pi^0}{\sqrt2}+\frac{\eta}{\sqrt6}&\pi^+&K^+\\
\pi^-&-\frac{\pi^0}{\sqrt2}+\frac{\eta}{\sqrt6}&K^0\\
K^-&\overline{K}^0&-\frac{2}{\sqrt6}\eta
\end{array}\right),\qquad
B=\left(
\begin{array}{ccc}
\frac{\Sigma^0}{\sqrt2}+\frac{\Lambda}{\sqrt6}&\Sigma^+&p\\
\Sigma^-&-\frac{\Sigma^0}{\sqrt2}+\frac{\Lambda}{\sqrt6}&n\\
\Xi^-&\Xi^0&-\frac{2}{\sqrt6}\Lambda
\end{array}\right),
\end{eqnarray}
\begin{equation}
i{\cal D}_\mu T^\nu_{abc}=i\partial_\mu T^\nu_{abc}+(\Gamma_\mu)^d_a
T^\nu_{dbc} + (\Gamma_\mu)^d_b T^\nu_{adc} +(\Gamma_\mu)^d_c
T^\nu_{abd},
\end{equation}
and
\begin{eqnarray}
T_{111}=\Delta^{++}, \qquad T_{112}=\frac{\Delta^{+}}{\sqrt3}, \qquad T_{122}=\frac{\Delta^{0}}{\sqrt3}, \qquad T_{222}=\Delta^{-}, \qquad T_{113}=\frac{\Sigma^{*+}}{\sqrt3},\nonumber\\
T_{123}=\frac{\Sigma^{*0}}{\sqrt6}, \qquad
T_{223}=\frac{\Sigma^{*-}}{\sqrt3}, \qquad
T_{133}=\frac{\Xi^{*0}}{\sqrt3}, \qquad
T_{233}=\frac{\Xi^{*-}}{\sqrt3}, \qquad T_{333}=\Omega^-.
\end{eqnarray}
$f$ is the pseudoscalar meson decay constant in the chiral limit.
$\Gamma_\mu$ is the chiral connection which contains even numbers
of meson fields. $u_\mu$ contains odd numbers of meson fields.
$D+F=g_A=1.26$ where $g_A$ is the axial vector coupling constant.
The superscripts in these Lagrangians represent the order of the
small scale expansion.

In our calculation of decuplet contributions to threshold
pseudoscalar meson octet-baryon scattering $T$-matrices, we truncate
at ${\cal O}(\epsilon^3)$. In this case, ${\cal L}^{(2)}_{\phi B}$
does not contribute, which can be found in \cite{KN,MB}. Similarly,
${\cal L}^{(2)}_{\phi BT}$ or high order Lagrangians also have
vanishing contributions.

Recently, the complete three-flavor Lorentz-invariant meson-baryon
chiral Lagrangians have been composed to the third order
\cite{Lp3,Lp32,Lp33}. Only three independent terms will contribute
to the meson-baryon scattering $T$-matrices at threshold
\begin{equation} \label{lagp3}
{\cal L}^{(3)}_{\phi B}=h_1{\rm tr}(\overline{B}B[\chi_-,u_0])+h_2
{\rm tr}(\overline{B}[\chi_-,u_0]B)+h_3 \{{\rm
tr}(\overline{B}u_0){\rm tr}(\chi_-B)-{\rm
tr}(\overline{B}\chi_-){\rm tr}(u_0B)\},
\end{equation}
where $h_1$, $h_2$ and $h_3$ are LECs, which also play the role of
absorbing divergences from loop calculations. When transforming the
relativistic Lagrangian into the heavy baryon formalism, additional
$1/M_0$ corrections may in principle appear. However, these kinds of
recoil corrections are higher than our truncation order. Thus the
above Lagrangian may also be treated as the form in HB$\chi$PT.

\section{Decuplet contributions to threshold $T$-matrices} \label{sec3}

\begin{figure}
\begin{center}
\includegraphics{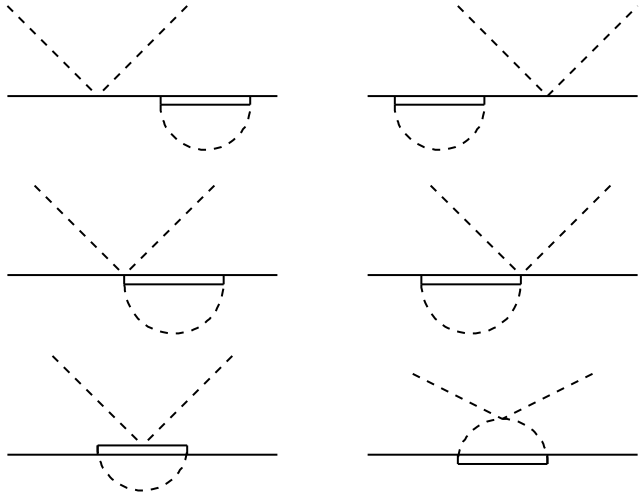}
\caption{Non-vanishing diagrams for meson-baryon scattering with
intermediate decuplet contributions at threshold. Dashed lines
represent Goldstone bosons, full lines represent octet baryons, and
double lines represent decuplet baryons.}\label{MBD}
\end{center}
\end{figure}

There are many diagrams for a general elastic pseudoscalar meson
octet-baryon scattering process to the third chiral order. When
intermediate decuplet contributions are considered, there are
additional diagrams. However, the calculation is simpler at
threshold. One may consult \cite{KN,MB} for the case with only
intermediate octet baryon. Here we consider intermediate decuplet
contributions. Decuplet corrections at the tree level vanish either
due to $\vec{\sigma}\cdot\vec{q}=0$ or $q^\mu P_{\mu\nu}^{3/2}=0$.
Here $\vec{\sigma}$ is the Pauli spin vector, $q$ is the momentum of
the external meson and $P_{\mu\nu}^{3/2}$ is the projection operator
of Rarita-Schwinger field. In the $d$-dimension space,
$P_{\mu\nu}^{3/2}=g_{\mu\nu}-v_\mu v_\nu+\frac{4}{d-1}S_\mu S_\nu$
where $g_{\mu\nu}$ is the metric tensor, $v_\mu$ is a four-velocity
and $S_\mu$ is the Pauli-Lubanski spin vector. Corrections to the
one-loop diagrams start from ${\cal O}(\epsilon^3)$ in the small
scale expansion. There are six non-vanishing diagrams at this order,
which we show in Fig. \ref{MBD}. The vertices in the figure are
generated from ${\cal L}^{(1)}_{\phi B}$, ${\cal L}^{(1)}_{\phi BT}$
and ${\cal L}^{(2)}_{\phi\phi}$.

In the previous loop calculations \cite{KN,MB}, dimensional
regularization and minimal subtraction were used. The divergences
were completely absorbed by the LECs in ${\cal L}_{\phi B}^{(3)}$.
When we consider the additional diagrams due to intermediate baryon
decuplet, there are divergences that the LECs will not absorb. We
give the finite parts from the loop calculation in this section and
give the counter-terms in the next section.

To write down the threshold $T$-matrices in compact forms, we define
\begin{eqnarray}
W(m^2)=\left\{
\begin{array}{ll}
\delta \ln\frac{|m|}{\lambda}
+\sqrt{\delta^2-m^2}\ln\frac{\delta+\sqrt{\delta^2-m^2}}{|m|}, & \mathrm{if} (m^2<\delta^2)\\
\delta \ln\frac{|m|}{\lambda}
-\sqrt{m^2-\delta^2}\arccos\frac{\delta}{|m|}, & \mathrm{if}
(m^2>\delta^2),
\end{array}
\right.
\end{eqnarray}
where $m$ represents the meson mass and $\lambda$ is the scale from
dimensional regularization. We list the $T$-matrices below.

For $\pi N$ scattering, we have
\begin{eqnarray}
T_{\pi N}^{(3/2)}=T_{\pi N}^{(1/2)}=-\frac{{\cal C}^2
m_\pi^2}{12\pi^2 f_\pi^4}\bigg\{W(m_\pi^2)\bigg\},
\end{eqnarray}
or
\begin{eqnarray}
T_{\pi N}^{+}=-\frac{{\cal C}^2 m_\pi^2}{12\pi^2
f_\pi^4}\bigg\{W(m_\pi^2)\bigg\},\qquad T_{\pi N}^{-}=0.
\end{eqnarray}

For $\pi\Sigma$ and $\pi\Xi$ scattering $T$-matrices, we have
\begin{eqnarray}
T_{\pi B}^{(I)}&=&\frac{{\cal C}^2 m_\pi^2}{48\pi^2 f_\pi^4}\bigg\{
\alpha_B^{(I)}W(m_\pi^2)+ W(m_\eta^2) \bigg\},
\end{eqnarray}
where $I$ labels the total isospin and $B$ the baryon $\Sigma$ or
$\Xi$,
\begin{eqnarray}
&\alpha_{\Sigma}^{(2)}=-1, \quad\alpha_{\Sigma}^{(1)}= 1,
\quad\alpha_{\Sigma}^{(0)}=-4;\qquad \alpha_{\Xi}^{(3/2)}=-1,
\qquad\alpha_{\Xi}^{(1/2)}=-1.&
\end{eqnarray}

The $T$-matrix for $\pi\Lambda$ scattering is very simple,
\begin{eqnarray}
T_{\pi \Lambda}&=&-\frac{{\cal C}^2m_\pi^2}{16\pi^2
f_\pi^4}\bigg\{W(m_\pi^2)\bigg\}.
\end{eqnarray}

The kaon-nucleon scattering $T$-matrices are
\begin{eqnarray}
T_{MN}^{(I)}&=&0,
\end{eqnarray}
where $M$ represents $K$ or $\overline{K}$ and $I=1$ or $I=0$.

The $T$-matrices for kaon-$\Sigma$ and kaon-$\Xi$ scatterings are
complicated due to the last diagram in Fig. \ref{MBD} with an
intermediate $\eta$ and an intermediate $\pi^0$. If we define
\begin{eqnarray}
J&=&\frac13\delta(1-6\ln\frac{m_\eta}{\lambda})-\frac{2\delta(3m_\pi^2-2\delta^2)}{3(m_\eta^2-m_\pi^2)}\ln\frac{m_\eta}{m_\pi}
\nonumber\\
&&+\frac{4}{3(m_\eta^2-m_\pi^2)}\bigg[(m_\eta^2-\delta^2)^{\frac32}\arccos{\frac{\delta}{
m_\eta}}-(\delta^2-m_\pi^2)^{\frac32}\ln\frac{\delta +
\sqrt{\delta^2 - m_\pi^2}}{m_\pi}\bigg],
\end{eqnarray}
the matrices for these two channels read:
\begin{eqnarray}
T_{MB}^{(I)}&=&\frac{{\cal C}^2 m_K^2}{48\pi^2
f_K^4}\bigg\{\alpha_{MB} ^{(I)}J -2 W(m_\eta^2) \bigg\},
\end{eqnarray}
where $M$ represents $K$ or $\overline{K}$, $B$ represents $\Sigma$
or $\Xi$,
\begin{eqnarray}
&\alpha_{K\Sigma}^{(3/2)}= 1,\quad\alpha_{K\Sigma}^{(1/2)}=-2,\quad
\alpha_{\overline{K}\Sigma}^{(3/2)}=-1,\alpha_{\overline{K}\Sigma}^{(1/2)}=2;&\nonumber\\
&\alpha_{K\Xi}^{(1)}=1,\quad\alpha_{K\Xi}^{(0)}=-3,
\quad\alpha_{\overline{K}\Xi}^{(1)}=-1,\quad\alpha_{\overline{K}\Xi}^{(0)}=1.&
\end{eqnarray}

The $T$-matrices for $T_{\overline{K}\Lambda}$ and $T_{K\Lambda}$
vanish.

The four $\eta B$ $T$-matrices depend on $W(m_\pi^2)$, $W(m_K^2)$
and $W(m_\eta^2)$.
\begin{eqnarray}
T_{\eta B}&=&\frac{{\cal C}^2}{144\pi^2 f_\eta^4}\bigg\{\alpha_B
m_\pi^2 W(m_\pi^2)+\beta_B m_K^2 W(m_K^2)+\gamma_B
(4m_\eta^2-m_\pi^2) W(m_\eta^2) \bigg\},
\end{eqnarray}
where
\begin{eqnarray}
&\alpha_N=12,\quad\beta_N=-6,\quad\gamma_N=0;\qquad
\alpha_\Sigma=2,\quad\beta_\Sigma=-20,\quad\gamma_\Sigma=3;&\nonumber\\
&\alpha_\Xi=3,\quad\beta_\Xi=-18,\quad\gamma_\Xi=3;\qquad
\alpha_\Lambda=9,\quad\beta_\Lambda=-12,\quad\gamma_\Lambda=0.&
\end{eqnarray}

In these loop expressions, we have used $f_\pi$ in pion processes,
$f_K$ in kaon processes and $f_\eta$ in eta processes. Different
usage of decay constants leads to deviations at higher order.
Therefore the difference in the numerical results is expected to
be negligible.

From these $T$-matrices, we see the intermediate decuplet states do
not generate any corrections for the imaginary part of the threshold
$T$-matrices. One can verify that the kaon-baryon and
anti-kaon-baryon $T$-matrices satisfy the crossing symmetry
\cite{KN,MB}. In the SU(3) limit, the relations in \cite{MB} also
hold. However, the similarity for $T$-matrices involving isospin
doublets do not exist any longer.

\section{Counter-terms}\label{sec4}

In the previous calculation of the threshold $T$-matrices, the
contributions from the renormalized counter-terms were naively
assumed to be much smaller than chiral loop corrections and ignored
in the numerical analysis. With the complete third-order Lagrangian
${\cal L}_{\phi B}^{(3)}$ \cite{Lp3,Lp32,Lp33}, we need to include
the counter-terms explicitly at ${\cal O}(\epsilon^3)$ from Eq.
(\ref{lagp3}).
\begin{equation}
T_{\pi N}^{(3/2)}=-4h^r_2\frac{m_\pi^3}{f_\pi^2},\qquad T_{\pi
N}^{(1/2)}=8h^r_2\frac{m_\pi^3}{f_\pi^2},\qquad T_{\pi
N}^{+}=0,\qquad T_{\pi N}^{-}=4h^r_2\frac{m_\pi^3}{f_\pi^2};
\end{equation}
\begin{eqnarray}
T_{\pi \Sigma}^{(2)}=4(h^r_1-h^r_2+h_3)\frac{m_\pi^3}{f_\pi^2},\quad
T_{\pi
\Sigma}^{(1)}=-4(h^r_1-h^r_2+h_3)\frac{m_\pi^3}{f_\pi^2},\quad
T_{\pi \Sigma}^{(0)}=-8(h^r_1-h^r_2+h_3)\frac{m_\pi^3}{f_\pi^2};
\end{eqnarray}
\begin{eqnarray}
T_{\pi \Xi}^{(3/2)}=4h^r_1\frac{m_\pi^3}{f_\pi^2},\qquad T_{\pi
\Xi}^{(1/2)}=-8h^r_1\frac{m_\pi^3}{f_\pi^2};
\end{eqnarray}
\begin{eqnarray}
&T_{KN}^{(1)}=4(h^r_1-h^r_2+h_3)\frac{m_K^3}{f_K^2},\quad
T_{KN}^{(0)}=4(h^r_1+h^r_2-h_3)\frac{m_K^3}{f_K^2},&\nonumber\\
&T_{\overline{K}N}^{(1)}=-4h^r_1\frac{m_K^3}{f_K^2},\quad
T_{\overline{K}N}^{(0)}=-4(h^r_1-2h^r_2+2h_3)\frac{m_K^3}{f_K^2};&
\end{eqnarray}
\begin{eqnarray}
&T_{K\Sigma}^{(3/2)}=-4h^r_2\frac{m_K^3}{f_K^2},\quad
T_{K\Sigma}^{(1/2)}=-2(3h^r_1-h^r_2)\frac{m_K^3}{f_K^2},&\nonumber\\
&T_{\overline{K}\Sigma}^{(3/2)}=4h^r_1\frac{m_K^3}{f_K^2},\quad
T_{\overline{K}\Sigma}^{(1/2)}=-2(h^r_1-3h^r_2)\frac{m_K^3}{f_K^2};&
\end{eqnarray}
\begin{eqnarray}
&T_{K\Xi}^{(1)}=4h^r_2\frac{m_K^3}{f_K^2},\quad
T_{K\Xi}^{(0)}=-4(2h^r_1-h^r_2+2h_3)\frac{m_K^3}{f_K^2},&\nonumber\\
&T_{\overline{K}\Xi}^{(1)}=4(h^r_1-h^r_2+h_3)\frac{m_K^3}{f_K^2},\quad
T_{\overline{K}\Xi}^{(0)}=-4(h^r_1+h^r_2+h_3)\frac{m_K^3}{f_K^2};&
\end{eqnarray}
\begin{eqnarray}
T_{\overline{K}\Lambda}=-T_{K\Lambda}=-2(h^r_1+h^r_2)\frac{m_K^3}{f_K^2};
\end{eqnarray}
\begin{equation}
T_{\pi \Lambda}=T_{\eta N}= T_{\eta\Sigma}= T_{\eta\Xi}= T_{\eta
\Lambda}=0.
\end{equation}

Among these counter-terms, $h_1^r$ and $h_2^r$ are renormalized LECs
while $h_3$ is finite. The renormalized LECs connect with
unrenormalized ones through $h_1^r=h_1+\frac34 \frac{L}{f^2}$ and
$h_2^r=h_2-\frac34 \frac{L}{f^2}$ when decuplet was not included.
$L=\frac{\lambda^{d-4}}{16\pi^2}[\frac{1}{d-4}+\frac12(\gamma_E-1-\ln
{4\pi})]$ where $d$ is the space-time dimension and $\gamma_E$ is
the Euler constant. When decuplet contributions are considered,
loops generate divergences proportional to $\delta m^2\frac{{\cal
C}^2 L}{f^4}$ which can not be cancelled. This is the result of
incompleteness for the renormalization in HB$\chi$PT with
$J=\frac32$ states because of non-vanishing $\delta$. One notes this
term is of ${1\over f^4}$ and analytical both in $\delta$ and $m$.
In the large $N_c$ limit, $\delta\rightarrow 0$ \cite{nc,ncWitten},
the correspondence between divergences and counter-terms recovers.
One practical approach is to keep non-analytical chiral corrections
only in the numerical analysis and throw away the above divergent
term with analytical coefficients.

One notes that the LECs $h_1^r$ and $h_2^r$ are in fact
scale-dependent. They cancel the scale-dependent parts arising from
loops, which makes the resulting $T$-matrices scale-independent. One
may define scale-independent LECs when only baryon octet is
involved. However, when decuplet is considered, one can not find
self-consistent definitions of scale-independent LECs. One can
verify these counter-terms satisfy the crossing symmetry and SU(3)
limit relations in \cite{MB}.

\section{Low-energy constants}\label{sec5}

To calculate the scattering lengths numerically, one has to
determine the LECs and their combinations. There are eight and three
parameters in the second and third-order Lagrangians, respectively.
The final $T$-matrices to ${\cal O}(\epsilon^3)$ involve five LECs
$b_D$, $b_F$, $h_1^r$, $h_2^r$, $h_3$ and four LEC combinations
$C_{1,0,\pi,d}$ which were defined as \cite{KN,MB}
\begin{eqnarray}
&C_1=2(d_0-2b_0)+2(d_D-2b_D)+d_1-{D^2+3F^2\over6M_0},&\nonumber\\
&C_0=2(d_0-2b_0)-2(d_F-2b_F)-d_1-\frac{D(D-3F)}{3M_0},&\nonumber\\
&C_\pi=(d_F-2b_F)-\frac{DF}{2M_0}, \qquad
C_d=d_1-\frac{D^2-3F^2}{6M_0} ,&
\end{eqnarray}
where $b_0$, $b_D$, $b_F$, $d_0$, $d_D$, $d_F$ and $d_1$ come from
${\cal L}_{\phi B}^{(2)}$. It is impossible to combine the LECs
$h_{1,2,3}$ with $C_{1,0,\pi,d}$ to reduce the number of
parameters. Up to now, we can determine most of them with
available inputs. Unfortunately, one is unable to extract $C_d$
strictly from known sources. We simply estimate its value.

In the SU(2) case, numerical evaluations for observables with either
scale-dependent or scale-independent LECs give the same results. In
the SU(3) case,  symmetry is largely broken. The two usages of LECs
may result in some deviations for current $T$-matrices, especially
when baryon decuplet contributions are considered. To reduce the
effects from symmetry breaking, we use scale-dependent $h_1^r$ and
$h_2^r$ in the following calculations. This choice is also
convenient for the discussion about the assumption used in
\cite{KN,MB}.

When determining the LECs, our procedure is as follows. (i) We first
choose the scale $\lambda=4\pi f_\pi$, which is widely used in
chiral perturbation theory. With the six pion-nucleon and
kaon-nucleon scattering lengths we determine $C_{1,0,\pi}$ and
$h_1^r, h_2^r, h_3$; (ii) then we determine $M_0, b_0, b_D, b_F$ by
fitting the baryon masses and $\sigma_{\pi N}$; (iii) finally, we
use these parameters as inputs to estimate other LECs and $C_d$. The
errors will also be estimated with the error propagation formula. We
consider three cases: the case with only baryon octet contributions,
with decuplet contributions and in the large $N_c$ limit.

Few experimental scattering lengths are available. Recently, those
for $\pi N$ scattering were measured \cite{TpiN}: $a_{\pi
N}^{+}=-0.0001^{+0.0009}_{-0.0021} \,m_\pi^{-1}, a_{\pi
N}^{-}=0.0885^{+0.0021}_{-0.0010} \,m_\pi^{-1}$. The new datum for
$a_{K^-p}$ \cite{DEAR} is not enough for our purpose. We use the
empirical values for kaon-nucleon scattering lengths from
\cite{TKN},
\begin{eqnarray}\label{emp}
a_{KN}^{(1)}=-0.33 \,{\rm fm}, \quad a_{KN}^{(0)}=0.02 \,{\rm fm},
\quad a_{\overline{K}N}^{(1)}=0.37+ 0.60 i \,{\rm fm}, \quad
a_{\overline{K}N}^{(0)}=-1.70+0.68 i \,{\rm fm}.
\end{eqnarray}

For the parameters in the expressions of $T$-matrices, we use
\cite{pdg}
\begin{eqnarray}\label{para}
&m_\pi=139.57 \,\mathrm{MeV},\quad m_K=493.68 \,\mathrm{MeV},\quad
m_\eta =547.75 \,\mathrm{MeV},&\nonumber\\
&\delta=294 \,\mathrm{MeV},\quad f_\pi=92.4 \,\mathrm{MeV},\quad f_K=113 \,\mathrm{MeV},\quad f_\eta=1.2 f_K&\nonumber\\
&D=0.75,\qquad F=0.5, \qquad {\cal C}=-1.5 \,\,.&
\end{eqnarray}

Now we reconsider the case with only baryon octet contributions by
including the counter-terms. The loop expressions can be found in
 \cite{MB}. When we express $C_{1,0,\pi}$, $h_1^r$, $h_2^r$ and
$h_3$ with $a_{\pi N}^+$, $a_{\pi N}^-$, $a_{KN}^{(1)}$,
$a_{KN}^{(0)}$, ${\rm Re}[a_{\overline{K}N}^{(1)}]$ and ${\rm
Re}[a_{\overline{K}N}^{(0)}]$, we get
\begin{eqnarray}
&C_1=-2.339 \,{\rm GeV^{-1}}, \quad C_0=4.389 \,{\rm GeV^{-1}},
\quad C_\pi=0.152^{+0.020}_{-0.048} \,{\rm GeV^{-1}},&\nonumber\\
&h_1^r=0.037 \,{\rm GeV^{-2}},\quad h_2^r=-0.274^{+0.171}_{-0.081}
\,{\rm GeV^{-2}}, \quad h_3=1.769^{+0.171}_{-0.081} \,{\rm
GeV^{-2}}.&
\end{eqnarray}

With the mass formulas in \cite{review,Dsigmas}, we get $b_D$,
$b_F$, $b_0$ and $M_0$ by fitting baryon masses $M_N=938.9\pm 1.3$
MeV, $M_\Sigma=1193.4\pm 8.1$ MeV, $M_\Xi=1318.1 \pm 6.7$ MeV,
$M_\Lambda=1115.7\pm 5.4$ MeV and $\sigma_{\pi N}=45 \pm 8
\,\mathrm{MeV}$ \cite{gasser},
\begin{eqnarray}
M_0=808.94\pm 104.20 \,\mathrm{MeV}, \qquad  b_0=-0.786\pm 0.103
\,\mathrm{GeV}^{-1}, \nonumber\\ b_D=0.028\pm 0.008
\,\mathrm{GeV}^{-1}, \qquad b_F=-0.473\pm 0.003 \,\mathrm{GeV}^{-1},
\end{eqnarray}
with $\chi^2/{\rm d.o.f.} \simeq0.75$. In the fitting procedure, we
have used $f=f_\pi$ in $\pi$ loops, $f=f_K$ in kaon loops and
$f=f_\eta$ in $\eta$ loops in the formulas. The results differ
slightly from our previous values only, because we used a smaller
$D$.

From the above determined quantities, we deduce
$d_F=-0.562^{+0.037}_{-0.057}\, {\rm GeV^{-1}}$ with $d_F=C_\pi +
2b_F +\frac{DF}{2M_0}$. Similarly, if we use the second order
$d_0=-0.996\, {\rm GeV^{-1}}$ \cite{d0}, we have
\begin{eqnarray}
d_D=0.331^{+0.413}_{-0.415}\, {\rm GeV^{-1}}, \quad
d_1=-3.772^{+0.414}_{-0.423}\, {\rm GeV^{-1}}, \quad
C_d=-3.733^{+0.414}_{-0.423}\, {\rm GeV^{-1}}.
\end{eqnarray}
One notes they are estimated values because of lack of
experimental inputs.

With intermediate decuplet contributions, one should note there are
still divergent parts in the $T$-matrices which could not be
absorbed. We just ignore them in the numerical evaluation as usually
done. By repeating the above procedure, we get
\begin{eqnarray}
&C_1=-2.339 \,{\rm GeV^{-1}}, \quad C_0=4.389 \,{\rm GeV^{-1}},
\quad
C_\pi=-0.145^{+0.020}_{-0.048} \,{\rm GeV^{-1}},&\nonumber\\
&h_1^r=0.037 \,{\rm GeV^{-2}}, \quad h_2^r=-0.274^{+0.171}_{-0.081}
\,{\rm GeV^{-2}}, \quad h_3=1.769^{+0.171}_{-0.081} \,{\rm
GeV^{-2}},&
\end{eqnarray}
with the updated threshold $T$-matrices.

When we determine $b_0,b_D,b_F$ and $M_0$, decuplet corrections to
baryon masses and $\pi N$ sigma term have to be considered. One gets
the corrections from the self energy diagram. If we define
\begin{equation}
I(m^2)=(m^2-\delta^2)W(m^2)-\delta(\frac14m^2-\frac13\delta^2)+\frac12\delta
m^2\ln\frac m\lambda,
\end{equation}
then the shifts of baryon masses are
\begin{eqnarray}
\Delta M_B&=&\frac{{\cal C}^2}{24\pi^2}\bigg\{\alpha_B
\frac{I(m_\pi^2)}{f_\pi^2} +\beta_B \frac{I(m_K^2)}{f_K^2} +\gamma_B
\frac{I(m_\eta^2)}{f_\eta^2}\bigg\},
\end{eqnarray}
where
\begin{eqnarray}
&\alpha_N=4,\quad\beta_N=1,\quad\gamma_N=0;\qquad\alpha_\Sigma=
\frac23,\quad\beta_\Sigma=\frac{10}{3},\quad\gamma_\Sigma=1;&\nonumber\\
&\alpha_\Xi=1,\quad\beta_\Xi=3,\quad\gamma_\Xi=1;\qquad\alpha_\Lambda=
3,\quad\beta_\Lambda=2,\quad\gamma_\Lambda=0.&
\end{eqnarray}
The shift of $\pi N$ $\sigma$ term is
\begin{equation}
\Delta\sigma_{\pi N}=\frac{m_\pi^2 {\cal
C}^2}{32\pi^2}\bigg\{8\frac{W(m_\pi^2)}{f_\pi^2}+\frac{W(m_K^2)}{f_K^2}\bigg\}.
\end{equation}
Here we give the finite parts only. In contrast to the case without
decuplet corrections, the divergent parts are non-vanishing. Those
proportional to $\delta m^2\frac{{\cal C}^2 L}{f^2}$ vanish either
in chiral limit or in large $N_c$ limit. Those proportional to
$\delta^3\frac{{\cal C}^2 L}{f^2}$ in baryon masses vanish only in
large $N_c$ limit. In \cite{Dsigmas}, a counter-term was added by
hand to cancel the latter divergence. In our case, we simply ignore
these regular parts. Our formulae are slightly different from those
in \cite{Dsigmas}.

With the updated mass and $\sigma_{\pi N}$ formulas, we get
\begin{eqnarray}
&M_0=745.02\pm 104.22 \,{\rm MeV}, \quad b_0=-1.342\pm 0.103
\,{\rm GeV^{-1}},&\nonumber\\
&b_D=0.308\pm 0.008 \,{\rm GeV^{-1}}, \quad b_F=-0.705\pm 0.003
\,{\rm GeV^{-1}}&
\end{eqnarray}
with $\chi^2/{\rm d.o.f.}\simeq 0.39$, which is smaller than the
former case. In this procedure, we again used $f_\pi$ in pion loops,
$f_K$ in kaon loops and $f_\eta$ in $\eta$ loops. Further we obtain
$d_F=-1.303^{+0.041}_{-0.060} \,{\rm GeV^{-1}}$,
$d_D=-1.608^{+0.414}_{-0.416} \,{\rm GeV^{-1}}$,
$d_1=-0.975^{+0.415}_{-0.423} \,{\rm GeV^{-1}}$ and
$C_d=-0.933^{+0.414}_{-0.423} \,{\rm GeV^{-1}}$ when we express them
with $C_{1,0,\pi}$, $b_0$, $b_D$, $M_0$ and $d_0$ and use $d_0=
-0.996 \,\mathrm{GeV}^{-1}$ \cite{d0} as inputs.

An important case is the large $N_c$ limit. In this limit,
$\delta=0, {\cal C}=-2D$ \cite{largeN}. The divergent parts related
with $\delta$ in $T$-matrices, baryon masses and $\pi N$ sigma term
vanish. The defined $W$ and $J$ are simple:
\begin{eqnarray}
W(m^2)=-\frac12 \pi |m|,  \qquad J=\frac{2\pi(m_\eta^2+m_\pi^2+m_\pi
m_\eta)}{3(m_\eta+m_\pi)}.
\end{eqnarray}
We also determine all the other LECs with $d_0 = -0.996
\,\rm{GeV}^{-1}$ in this limit by repeating the above procedure.
They are
\begin{eqnarray}
&M_0=808.94\pm 104.20 \,{\rm MeV}, \quad b_0=-0.786\pm 0.103
\,{\rm GeV^{-1}}, \quad b_D=0.028\pm 0.008 \,{\rm GeV^{-1}},&\nonumber\\
& b_F=-0.473\pm 0.003 \,{\rm GeV^{-1}},\quad
d_F=-0.806^{+0.037}_{-0.057} \,{\rm GeV^{-1}},&\nonumber\\
&d_D=0.087^{+0.413}_{-0.415} \,{\rm GeV^{-1}}, \quad
d_1=-3.284^{+0.414}_{-0.423} \,{\rm GeV^{-1}},& \nonumber\\
&h_1^r=-0.037 \,{\rm GeV^{-2}}, \quad h_2^r=-0.274^{+0.171}_{-0.081}
\,{\rm GeV^{-2}}, \quad h_3=1.769^{+0.171}_{-0.081} \,{\rm
GeV^{-2}}.&
\end{eqnarray}
When we obtain the first four parameters with MINUIT, we have
$\chi^2/{\rm d.o.f.}\simeq 0.75$. Those LEC combinations are
\begin{eqnarray}
&C_1= -2.339 \,\mathrm{GeV}^{-1}, \quad C_0= 4.389
\,\mathrm{GeV}^{-1}, \quad C_\pi=-0.092^{+0.020}_{-0.048}
\,\mathrm{GeV}^{-1},\quad C_d=-3.245^{+0.414}_{-0.423}
\,\mathrm{GeV}^{-1}.&
\end{eqnarray}

\section{Numerical results and discussions}\label{sec6}

With the three sets of parameters, we evaluate all the meson-baryon
threshold $T$-matrices. We present numerical results involving only
octet baryons in Tables \ref{piB}, \ref{kaonB} and \ref{etaB}, the
results for the case with decuplet contributions in Tables
\ref{piBD}, \ref{kaonBD} and \ref{etaBD} and those in the large
$N_c$ limit in Tables \ref{piBN}, \ref{kaonBN} and \ref{etaBN}. The
corresponding scattering lengths are given in the last column. We
estimate the errors from those of $C_\pi$, $C_d$, $b_D$, $b_F$,
$h_2^r$ and $h_3$ with the error propagation formula.

\begin{table}
\caption{Pion-baryon threshold $T$-matrices order by order with only
octet contributions in unit of fm. }\label{piB}
\begin{tabular}{cccccc}
\hline
                      &${\cal O}(p)$&${\cal O}(p^2)$&${\cal O}(p^3)$&Total&Scattering lengths\\\hline
$T_{\pi N}^+$         & 0           & $0.60^{+0.02}_{-0.04}$          & $-0.60$         & $-0.002^{+0.018}_{-0.043}$  &$-0.00014^{+0.00127}_{-0.00297}$ (input)  \\
$T_{\pi N}^-$        & 1.61         & 0             & $0.19^{+0.04}_{-0.02}$          & $1.81^{+0.04}_{-0.02}$     &$0.125^{+0.003}_{-0.001}$ (input)  \\
$T_{\pi N}^{(3/2)}$  & $-1.61$      & $0.60^{+0.02}_{-0.04}$         & $-0.79^{+0.04}_{-0.02}$         & $-1.81\pm 0.05$   &$-0.130\pm 0.003$   \\
$T_{\pi N}^{(1/2)}$  & 3.23         & $0.60^{+0.02}_{-0.04}$         & $-0.21^{+0.08}_{-0.04}$        & $3.61^{+0.18}_{-0.12}$   &$0.25\pm 0.01$   \\
$T_{\pi\Sigma}^{(2)}$& $-3.23$        &$-1.05$           &$-0.50^{+0.06}_{-0.03}$          &$-4.77^{+0.06}_{-0.03}$     &$-0.340^{+0.004}_{-0.002}$\\
$T_{\pi\Sigma}^{(1)}$&  3.23        &$2.31^{+0.37}_{-0.38}$          &$-0.55^{+0.06}_{-0.03}$         & $4.99\pm0.38$   &$0.36\pm{0.03}$  \\
$T_{\pi\Sigma}^{(0)}$&  6.45        &$-6.09^{+0.56}_{-0.57}$          & $-0.41^{+0.12}_{-0.06}$          & $-0.05\pm0.57$   &$-0.004\pm{0.041}$  \\
$T_{\pi\Xi}^{(3/2)}$ & $-1.61$        & 0.46          & $-1.23$         & $-2.38$    &$-0.17$       \\
$T_{\pi\Xi}^{(1/2)}$ & 3.23         & 0.46          & $-0.48$         & 3.21     &0.23        \\
$T_{\pi\Lambda}$     & 0            & $0.50\pm0.06$          & $-1.52$         & $-1.02\pm0.06$   &$-0.072\pm0.004 $ \\
\hline
\end{tabular}
\end{table}

\begin{table}
\caption{Kaon-baryon threshold $T$-matrices order by order with only
octet contributions in unit of fm. }\label{kaonB}
\begin{tabular}{cccccc}
\hline
                      &${\cal O}(p)$&${\cal O}(p^2)$&${\cal O}(p^3)$&Total&Scattering lengths\\\hline
$T_{KN}^{(1)}$        & $-7.63$       &  $-8.81$         & $10.11$         & $-6.33$          &$-0.33$ (input) \\
$T_{KN}^{(0)}$        & 0           & 16.53          & $-16.15$         & 0.38           &0.02 (input)  \\
$T_{\overline{K}N}^{(1)}$  & 3.81        & 3.86          & $-0.58+6.95i$   & $7.09+6.95i$     &$0.37+0.36i$ (input) \\
$T_{\overline{K}N}^{(0)}$  & 11.44       &$-21.48$           & $-22.56+4.17i$    & $-32.60+4.17i$    &$-1.70+0.22i$ (input)\\
$T_{K\Sigma}^{(3/2)}$ & $-3.81$        &$5.01^{+0.15}_{-0.36}$    &$0.75^{+1.27}_{-0.60}+2.78 i$   &$1.95^{+1.28}_{-0.70}+2.78 i$   &$0.11^{+0.07}_{-0.04}+0.16i$ \\
$T_{K\Sigma}^{(1/2)}$ &  7.63        &$3.29^{+0.08}_{-0.18}$    &$1.62^{+0.63}_{-0.30}+0.69i$     &$12.54^{+0.64}_{-0.35}+0.69i$    &$0.71^{+0.04}_{-0.02}+0.04i$ \\
$T_{\overline{K}\Sigma}^{(3/2)}$& $-3.81$       &3.86           &$-4.39+2.78i$  &$-4.34+2.78i$   &$-0.24+0.16i$   \\
$T_{\overline{K}\Sigma}^{(1/2)}$& 7.63        &$5.58^{+0.23}_{-0.54}$           &$6.62^{+1.90}_{-0.91}+0.69i$   &$19.83^{+1.92}_{-1.05}+0.69i$   &$1.12^{+0.11}_{-0.06}+0.04i$   \\
$T_{K\Xi}^{(1)}$      & 3.81        & $5.01^{+0.15}_{-0.36}$          & $1.76^{+1.27}_{-0.60}+6.95i$    & $10.58^{+1.28}_{-0.70}+6.95i$    &$0.61^{+0.07}_{-0.04}+0.40i$  \\
$T_{K\Xi}^{(0)}$      &11.44        & $-22.62^{+0.15}_{-0.36}$          & $-23.46^{+2.84}_{-1.35}+4.17i$    & $-34.64^{+2.84}_{-1.40}+4.17i$    &$-2.01^{+0.16}_{-0.08}+0.24i$  \\
$T_{\overline{K}\Xi}^{(1)}$& $-7.63$       & $-8.81$          & $10.83^{+1.79}_{-0.85}$         & $-5.61^{+1.79}_{-0.85}$          &$-0.32^{+0.10}_{-0.05}$  \\
$T_{\overline{K}\Xi}^{(0)}$& 0           & $18.82^{+0.31}_{-0.72}$          & $-5.14^{+1.79}_{-0.85}$          & $13.68^{+1.82}_{-1.12}$          &$0.79^{+0.11}_{-0.06}$   \\
$T_{K\Lambda}           $& 0   & $4.98^{+1.04}_{-1.08}$          & $-2.81^{+0.63}_{-0.30}+6.25i$   & $2.17^{+1.22}_{-1.12}+6.25i$  &$0.12^{+0.07}_{-0.06}+0.34i$  \\
$T_{\overline{K}\Lambda}$& 0   & $4.98^{+1.04}_{-1.08}$          & $-1.05^{+0.63}_{-0.30}+6.25i$   & $3.93^{+1.22}_{-1.12}+6.25i$  &$0.22^{+0.07}_{-0.06}+0.34i$  \\
\hline
\end{tabular}
\end{table}

\begin{table}
\caption{Eta-baryon threshold $T$-matrices order by order with only
octet contributions in unit of fm. }\label{etaB}
\begin{tabular}{ccccc}
\hline
                  &${\cal O}(p^2)$&${\cal O}(p^3)$&Total     &Scattering lengths\\\hline
$T_{\eta N}$          & $1.29^{+1.32}_{-1.33}$      & $2.06+8.32i$   & $3.35^{+1.32}_{-1.33}+8.32i$  &$(0.17\pm 0.07)+0.42i$\\
$T_{\eta\Sigma}$      & $5.49^{+0.67}_{-0.68}$      & $1.88+5.55i$   & $7.37^{+0.67}_{-0.68}+5.55i$   &$(0.40\pm0.04)+0.30i$ \\
$T_{\eta\Xi}$         & $11.28^{+1.35}_{-1.46}$     & $0.53+8.32i$   &$11.80^{+1.35}_{-1.46}+8.32i$  &$(0.66\pm0.08)+0.47i$\\
$T_{\eta\Lambda}$     & $-29.41^{+1.99}_{-2.04}$     & $2.67+16.64i$  & $-26.73^{+1.99}_{-2.04}+16.64i$ &$(-1.43\pm 0.11)+0.89i$\\
\hline
\end{tabular}
\end{table}

\begin{table}
\caption{Pion-baryon threshold $T$-matrices order by order,
including decuplet contributions in unit of fm. }\label{piBD}
\begin{tabular}{cccccc}
\hline
                      &${\cal O}(\epsilon)$&${\cal O}(\epsilon^2)$&${\cal O}(\epsilon^3)$&Total&Scattering lengths\\\hline
$T_{\pi N}^+$         & 0           & $0.33^{+0.02}_{-0.04}$          & $-0.33$         & $-0.002^{+0.018}_{-0.043}$  &$-0.00014^{+0.00127}_{-0.00297}$ (input)  \\
$T_{\pi N}^-$        & 1.61         & 0             & $0.19^{+0.04}_{-0.02}$          & $1.81^{+0.04}_{-0.02}$     &$0.125^{+0.003}_{-0.001}$ (input) \\
$T_{\pi N}^{(3/2)}$  & $-1.61$        & $0.33^{+0.02}_{-0.04}$         & $-0.53^{+0.04}_{-0.02}$         & $-1.81\pm0.05$   &$-0.125\pm0.003$   \\
$T_{\pi N}^{(1/2)}$  & 3.23         & $0.33^{+0.02}_{-0.04}$         & $0.05^{+0.09}_{-0.04}$        & $3.61^{+0.18}_{-0.12}$   &$0.25\pm0.01$   \\
$T_{\pi\Sigma}^{(2)}$& $-3.23$        &$-1.05$           &$0.60^{+0.06}_{-0.03}$          &$-4.88^{+0.06}_{-0.03}$     &$-0.35^{+0.004}_{-0.002}$\\
$T_{\pi\Sigma}^{(1)}$&  3.23        &$-0.21^{+0.37}_{-0.38}$          &$-0.79^{+0.06}_{-0.03}$         & $2.23^{+0.37}_{-0.38}$   &$0.16\pm{0.03}$  \\
$T_{\pi\Sigma}^{(0)}$&  6.45        &$-2.31^{+0.56}_{-0.57}$          & $-0.31^{+0.12}_{-0.06}$          & $3.82\pm0.57$   &$0.27\pm{0.04}$  \\
$T_{\pi\Xi}^{(3/2)}$ & $-1.61$        & 0.46          & $-1.34$         & $-2.49$    &$-0.18$    \\
$T_{\pi\Xi}^{(1/2)}$ & 3.23         & 0.46          & $-0.58$         & $3.11$     &$0.22$   \\
$T_{\pi\Lambda}$     & 0            & $0.74\pm0.06$          & $-1.32$         & $-0.58\pm0.06$   &$-0.041\pm0.004 $ \\
\hline
\end{tabular}
\end{table}

\begin{table}
\caption{Kaon-baryon threshold $T$-matrices order by order,
including decuplet contributions in unit of fm. }\label{kaonBD}
\begin{tabular}{cccccc}
\hline
                      &${\cal O}(\epsilon)$&${\cal O}(\epsilon^2)$&${\cal O}(\epsilon^3)$&Total&Scattering lengths\\\hline
$T_{KN}^{(1)}$        & $-7.63$       &  $-8.81$         & $10.11$         & $-6.33$          &$-0.33$ (input) \\
$T_{KN}^{(0)}$        & 0           & 16.53          & $-16.15$         & 0.38           &0.02 (input)  \\
$T_{\overline{K}N}^{(1)}$  & 3.81        & 3.86          & $-0.58+6.95i$   & $7.09+6.95i$     &$0.37+0.36i$(input) \\
$T_{\overline{K}N}^{(0)}$  & 11.44       &$-21.48$           & $-22.56+4.17i$    & $-32.60+4.17i$    &$-1.70+0.22i$ (input)\\
$T_{K\Sigma}^{(3/2)}$ & $-3.81$        &$2.77^{+0.15}_{-0.36}$    &$3.90^{+1.27}_{-0.60}+2.78 i$   &$2.86^{+1.28}_{-0.70}+2.78 i$   &$0.16^{+0.07}_{-0.04}+0.16i$ \\
$T_{K\Sigma}^{(1/2)}$ &  7.63        &$4.41^{+0.08}_{-0.18}$    &$1.08^{+0.63}_{-0.30}+0.69i$     &$13.11^{+0.64}_{-0.35}+0.69i$    &$0.74^{+0.04}_{-0.02}+0.04i$ \\
$T_{\overline{K}\Sigma}^{(3/2)}$& $-3.81$       &3.86           &$-3.70+2.78i$  &$-3.65+2.78i$   &$-0.21+0.16i$   \\
$T_{\overline{K}\Sigma}^{(1/2)}$& 7.63        &$2.23^{+0.23}_{-0.54}$           &$11.00^{+1.90}_{-0.91}+0.69i$   &$20.86^{+1.92}_{-1.05}+0.69i$   &$1.17^{+0.11}_{-0.06}+0.04i$   \\
$T_{K\Xi}^{(1)}$      & 3.81        & $2.77^{+0.15}_{-0.36}$          & $4.91^{+1.27}_{-0.60}+6.95i$    & $11.49^{+1.28}_{-0.70}+6.95i$    &$0.67^{+0.07}_{-0.04}+0.40i$  \\
$T_{K\Xi}^{(0)}$      &11.44        & $-20.39^{+0.15}_{-0.36}$          & $-25.23^{+2.84}_{-1.35}+4.17i$    & $-34.18^{+2.84}_{-1.40}+4.17i$    &$-1.98^{+0.16}_{-0.08}+0.24i$  \\
$T_{\overline{K}\Xi}^{(1)}$& $-7.63$       & $-8.81$          & $11.52^{+1.79}_{-0.85}$         & $-4.92^{+1.79}_{-0.85}$          &$-0.28^{+0.10}_{-0.05}$  \\
$T_{\overline{K}\Xi}^{(0)}$& 0           & $14.35^{+0.31}_{-0.72}$          & $0.47^{+1.79}_{-0.85}$          & $14.82^{+1.82}_{-1.12}$          &$0.86^{+0.11}_{-0.06}$ \\
$T_{K\Lambda}           $& 0   & $-2.42^{+1.04}_{-1.08}$          & $-2.81^{+0.63}_{-0.30}+6.25i$   & $-5.23^{+1.22}_{-1.12}+6.25i$  &$-0.29^{+0.07}_{-0.06}+0.34i$  \\
$T_{\overline{K}\Lambda}$& 0   & $-2.42^{+1.04}_{-1.08}$          & $-1.05^{+0.63}_{-0.30}+6.25i$   &$-3.47^{+1.22}_{-1.12}+6.25i$  &$-0.19^{+0.07}_{-0.06}+0.34i$  \\
\hline
\end{tabular}
\end{table}

\begin{table}
\caption{Eta-baryon threshold $T$-matrices order by order, including
decuplet contributions in unit of fm. }\label{etaBD}
\begin{tabular}{ccccc}
\hline
                  &${\cal O}(\epsilon^2)$&${\cal O}(\epsilon^3)$&Total     &Scattering lengths\\\hline
$T_{\eta N}$          & $-9.67^{+1.32}_{-1.33}$      & $3.81+8.32i$   & $-5.87^{+1.32}_{-1.33}+8.32i$  &$(-0.29\pm 0.07)+0.42i$\\
$T_{\eta\Sigma}$      & $9.75^{+0.67}_{-0.68}$      & $3.10+5.55i$   & $12.86^{+0.67}_{-0.68}+5.55i$   &$(0.70\pm0.04)+0.30i$ \\
$T_{\eta\Xi}$         & $1.63^{+1.35}_{-1.46}$     & $1.12+8.32i$   &$2.75^{+1.35}_{-1.46}+8.32i$  &$(0.15\pm0.08)+0.47i$\\
$T_{\eta\Lambda}$     & $-17.60^{+1.99}_{-2.04}$     & $6.33+16.64i$  & $-11.27^{+1.99}_{-2.04}+16.64i$ &$(-0.60\pm 0.11)+0.89i$\\
\hline
\end{tabular}
\end{table}

\begin{table}
\caption{Pion-baryon threshold $T$-matrices order by order in large
$N_c$ limit in unit of fm. }\label{piBN}
\begin{tabular}{cccccc}
\hline
                      &${\cal O}(\epsilon)$&${\cal O}(\epsilon^2)$&${\cal O}(\epsilon^3)$&Total&Scattering lengths\\\hline
$T_{\pi N}^+$         & 0           & $0.38^{+0.02}_{-0.04}$          & $-0.38$         & $-0.002^{+0.018}_{-0.043}$  &$-0.00014^{+0.00127}_{-0.00297}$ (input)  \\
$T_{\pi N}^-$        & 1.61         & 0             & $0.19^{+0.04}_{-0.02}$          & $1.81^{+0.04}_{-0.02}$     &$0.125^{+0.003}_{-0.001}$ (input)  \\
$T_{\pi N}^{(3/2)}$  & $-1.61$      & $0.38^{+0.02}_{-0.04}$         & $-0.57^{+0.04}_{-0.02}$         & $-1.81\pm 0.05$   &$-0.125\pm 0.003$   \\
$T_{\pi N}^{(1/2)}$  & 3.23         & $0.38^{+0.02}_{-0.04}$         & $-0.01^{+0.09}_{-0.04}$        & $3.61^{+0.18}_{-0.12}$   &$0.25\pm 0.01$   \\
$T_{\pi\Sigma}^{(2)}$& $-3.23$        &$-1.05$           &$-0.66^{+0.06}_{-0.03}$          &$-4.94^{+0.06}_{-0.03}$     &$-0.35^{+0.004}_{-0.002}$\\
$T_{\pi\Sigma}^{(1)}$&  3.23        &$1.87^{+0.37}_{-0.38}$          &$-0.82^{+0.06}_{-0.03}$         & $4.28\pm0.38$   &$0.30\pm{0.03}$  \\
$T_{\pi\Sigma}^{(0)}$&  6.45        &$-5.44^{+0.56}_{-0.57}$          & $-0.41^{+0.12}_{-0.06}$          & $0.61\pm0.57$   &$0.04\pm{0.04}$  \\
$T_{\pi\Xi}^{(3/2)}$ & $-1.61$        & 0.46          & $-1.39$         & $-2.55$    &$-0.18$  \\
$T_{\pi\Xi}^{(1/2)}$ & 3.23         & 0.46          & $-0.64$         & 3.05     &0.22        \\
$T_{\pi\Lambda}$     & 0            & $0.42\pm0.06$          & $-1.35$         & $-0.93\pm0.06$   &$-0.066\pm0.004 $ \\
\hline
\end{tabular}
\end{table}

\begin{table}
\caption{Kaon-baryon threshold $T$-matrices order by order in large
$N_c$ limit in unit of fm. }\label{kaonBN}
\begin{tabular}{cccccc}
\hline
                      &${\cal O}(\epsilon)$&${\cal O}(\epsilon^2)$&${\cal O}(\epsilon^3)$&Total&Scattering lengths\\\hline
$T_{KN}^{(1)}$        & $-7.63$       &  $-8.81$         & $10.11$         & $-6.33$          &$-0.33$ (input) \\
$T_{KN}^{(0)}$        & 0           & 16.53          & $-16.15$         & 0.38           &0.02 (input)  \\
$T_{\overline{K}N}^{(1)}$  & 3.81        & 3.86          & $-0.58+6.95i$   & $7.09+6.95i$     &$0.37+0.36i$ (input) \\
$T_{\overline{K}N}^{(0)}$  & 11.44       &$-21.48$           & $-22.56+4.17i$    & $-32.60+4.17i$    &$-1.70+0.22i$ (input)\\
$T_{K\Sigma}^{(3/2)}$ & $-3.81$        &$3.17^{+0.15}_{-0.36}$    &$4.62^{+1.27}_{-0.60}+2.78 i$   &$3.98^{+1.28}_{-0.70}+2.78 i$   &$0.22^{+0.07}_{-0.04}+0.16i$ \\
$T_{K\Sigma}^{(1/2)}$ &  7.63        &$4.21^{+0.08}_{-0.18}$    &$1.11^{+0.63}_{-0.30}+0.69i$     &$12.95^{+0.64}_{-0.35}+0.69i$    &$0.73^{+0.04}_{-0.02}+0.04i$ \\
$T_{\overline{K}\Sigma}^{(3/2)}$& $-3.81$       &3.86           &$-3.44+2.78i$  &$-3.39+2.78i$   &$-0.19+0.16i$   \\
$T_{\overline{K}\Sigma}^{(1/2)}$& 7.63        &$2.82^{+0.23}_{-0.54}$           &$11.95^{+1.90}_{-0.91}+0.69i$   &$20.40^{+1.92}_{-1.05}+0.69i$   &$1.26^{+0.11}_{-0.06}+0.04i$   \\
$T_{K\Xi}^{(1)}$      & 3.81        & $3.17^{+0.15}_{-0.36}$          & $5.63^{+1.27}_{-0.60}+6.95i$    & $12.61^{+1.28}_{-0.70}+6.95i$    &$0.73^{+0.07}_{-0.04}+0.40i$  \\
$T_{K\Xi}^{(0)}$      &11.44        & $-20.79^{+0.15}_{-0.36}$          & $-25.43^{+2.84}_{-1.35}+4.17i$    & $-34.77^{+2.84}_{-1.40}+4.17i$    &$-2.01^{+0.16}_{-0.08}+0.24i$  \\
$T_{\overline{K}\Xi}^{(1)}$& $-7.63$       & $-8.81$          & $11.78^{+1.79}_{-0.85}$         & $-4.65^{+1.79}_{-0.85}$          &$-0.27^{+0.10}_{-0.05}$  \\
$T_{\overline{K}\Xi}^{(0)}$& 0           & $15.15^{+0.31}_{-0.72}$          & $1.65^{+1.79}_{-0.85}$          & $16.79^{+1.82}_{-1.12}$          &$0.97^{+0.11}_{-0.06}$   \\
$T_{K\Lambda}           $& 0   & $3.45^{+1.04}_{-1.08}$          & $-2.81^{+0.63}_{-0.30}+6.25i$   & $0.64^{+1.22}_{-1.12}+6.25i$  &$0.04^{+0.07}_{-0.06}+0.34i$  \\
$T_{\overline{K}\Lambda}$& 0   & $3.45^{+1.04}_{-1.08}$          & $-1.05^{+0.63}_{-0.30}+6.25i$   & $2.40^{+1.22}_{-1.12}+6.25i$  &$0.13^{+0.07}_{-0.06}+0.34i$  \\
\hline
\end{tabular}
\end{table}

\begin{table}
\caption{Eta-baryon threshold $T$-matrices order by order in large
$N_c$ limit in unit of fm. }\label{etaBN}
\begin{tabular}{ccccc}
\hline
                  &${\cal O}(\epsilon^2)$&${\cal O}(\epsilon^3)$&Total     &Scattering lengths\\\hline
$T_{\eta N}$          & $0.51^{+1.32}_{-1.33}$      & $4.30+8.32i$   & $4.80^{+1.32}_{-1.33}+8.32i$  &$(0.24\pm 0.07)+0.42i$\\
$T_{\eta\Sigma}$      & $4.70^{+0.67}_{-0.68}$      & $3.38+5.55i$   & $8.08^{+0.67}_{-0.68}+5.55i$   &$(0.44\pm0.04)+0.30i$ \\
$T_{\eta\Xi}$         & $8.14^{+1.35}_{-1.46}$     & $1.23+8.32i$   &$9.38^{+1.35}_{-1.46}+8.32i$  &$(0.53\pm0.08)+0.47i$\\
$T_{\eta\Lambda}$     & $-27.06^{+1.99}_{-2.04}$     & $7.28+16.64i$  & $-19.78^{+1.99}_{-2.04}+16.64i$ &$(-1.06\pm 0.11)+0.89i$\\
\hline
\end{tabular}
\end{table}

We first consider the case without explicit decuplet
contributions. In this case, the contributions from the decuplet
baryons, resonances close to thresholds and other baryons are all
buried in the LECs.

It is interesting to see those eta-baryon scattering lengths. The
widely studied $\eta$-mesic nuclei were proposed decades ago
\cite{etamesic,etamesic2}. The $\eta$-mesic hypernuclei were also
proposed \cite{etahy}. An important parameter to verify whether they
exist is the eta-baryon scattering length. One can consult \cite{MB}
for our previous calculation for this observable. When counter-terms
were considered, we find that ${\rm Re}[a_{\eta\Lambda}]$ is
negative, contrary to the results in \cite{etahy,I0MB}. This is due
to the large and negative contribution at the second order. From
$T_{\eta\Lambda}$ at the second order and the values $C_1$, $C_d$
and $b_D$, it is not surprising to get this result. One may think
that the estimated $C_d$ is problematic. According to $C_d=
-C_0-2C_\pi-4b_0+2d_0-\frac{D^2-F^2}{2M_0}$, we see that even if
$d_0\sim 0.0$, $C_d$ would still be negative. As a result, it is
almost definite that ${\rm Re}[a_{\eta\Lambda}] < 0$.

From the results in Tables \ref{piB}, \ref{kaonB} and \ref{etaB}, we
find that the $T$-matrices $T_{\pi N}^{(1/2)}$,
$T_{\pi\Sigma}^{(2)}$, $T_{\pi\Sigma}^{(1)}$ and
$T_{K\Sigma}^{(1/2)}$ converge well. The situation differs slightly
from the previous calculation \cite{MB}. Any scattering lengths
related with $C_d$ can be used as a constraint to determine LECs if
they were measured, but $a_{\pi\Sigma}^{(1)}$ is particularly ideal
for this purpose.

Let us discuss the contributions from counter-terms. The
pion-nucleon scattering lengths were first calculated to ${\cal
O}(p^3)$ in \cite{pnsl} in the SU(2) HB$\chi$PT. The counter-terms
were estimated with resonance saturation method. It was found that
counter-terms have much smaller contributions than chiral loop
corrections. This naive assumption was extended to the SU(3) case in
\cite{KN,MB}. However, such an extension is actually problematic. We
have analyzed the third order $T$-matrices numerically. We found
that counter-terms have even larger contributions than loops in the
following $T$-matrices: $T_{\pi\Sigma}^{(1)}$,
$T_{\pi\Sigma}^{(0)}$, $T_{KN}^{(1)}$, $T_{KN}^{(0)}$,
$T_{\overline{K}N}^{(0)}$, $T_{K\Sigma}^{(3/2)}$, $T_{K\Xi}^{(0)}$,
$T_{\overline{K}\Xi}^{(1)}$ and $T_{\overline{K}\Xi}^{(0)}$.
Fortunately, most of the predictions in \cite{MB} are not far away
from the current calculation.

When the baryon decuplet contributions were considered explicitly,
about half of the predictions are close to those without decuplet
contribution. But several scattering lengths change sign. Now the
real part of $a_{\eta N}$ has a negative sign which is contrary to
results in literature (see \cite{etaNcompi} for an overview).
$\mathrm{Re}(a_{\eta\Sigma})=0.70\pm 0.04$ fm is still consistent
with $a_{\eta\Sigma}=[(0.10\sim 1.10) +(0.35\sim 2.20)i]\,\rm{fm}$
in \cite{etahy}. ${\rm Re}[a_{\eta\Lambda}]$ is still negative. The
change of scattering lengths due to the inclusion of decuplet is
known to be the result of non-commutativity of the chiral limit and
large $N_c$ limit. The effects of the interplay of these two limits
on meson-baryon scattering lengths were recently discussed in
\cite{limeff}.

From Tables 4-6, 
one finds that the chiral expansion converges for $T_{\pi
N}^{(1/2)}$, $T_{\pi\Sigma}^{(2)}$, $T_{\pi\Sigma}^{(0)}$,
$T_{K\Sigma}^{(1/2)}$, $T_{K\Lambda}$, and
$T_{\overline{K}\Lambda}$. It was pointed out in \cite{axialD,vec}
that the chiral expansion in HB$\chi$PT with decuplet is worse than
that in HB$\chi$PT with only ground baryons. In the present case,
the convergence of scattering lengths changes little with decuplet
corrections.

All known scattering lengths have been used to determine LECs. There
are no available experimental data to compare with. The present
predictions require future experimental measurements to test
HB$\chi$PT. On the other hand, once the four kaon-nucleon scattering
lengths as well as $a_{\pi\Sigma}$ were measured accurately, we can
predict precisely others to the third order in HB$\chi$PT.
Strangeness programmes in CSR and JPARC are hopeful for these
purposes.

In summary, we have calculated intermediate decuplet baryon
contributions to the s-wave meson-baryon scattering lengths to the
third order in small scale expansion. Hopefully, the explicit
expressions are helpful to future chiral extrapolations in lattice
simulations. From the known $a_{\pi N}$, $a_{KN}$ and
$a_{\overline{K}N}$, we determined the LECs and predicted other
scattering lengths in three cases by including the counter-terms. We
found that chiral expansion in several channels converges well
without considering the decuplet contributions. When decuplet
contributions were considered, the convergence of the chiral
expansion does not change significantly. Our calculation indicates
that $a_{\eta\Lambda}$ is negative. Whether $\eta$-mesic hypernuclei
is possible requires further investigations. We expect the numerical
results are useful to model constructions for meson-baryon
interaction.

\section*{Acknowledgments}

This project was supported by the National Natural Science
Foundation of China under Grants 10421503 and 10625521, Ministry
of Education of China, FANEDD, and Key Grant Project of Chinese
Ministry of Education (NO 305001). Y.R.L. thanks Y. Cui for
checking part of the calculation.

\end{document}